# Beam Emittance Measurement for PLS-II Linac

**Byung-Joon Lee, Ilmoon Hwang, Chong do Park, Changbum Kim***

*Pohang Accelerator Laboratory, Pohang, Korea, 790-784*

**SomJai Chunjarean**

*Chiang Mai University, 239 Huay Kaew Road, Muang District, Chiang Mai, 50200, Thailand*

The PLS-II has a 100 MeV pre-injector for the 3 GeV Linac. A thermionic gun produces electron charge of 200 pC with a bunch duration of 500 ps by a 250 ps triggering pulser. At the pre-injector, one of the most important beam parameters to identify the beam quality is a transverse emittance of electron bunches. Therefore we measure the beam emittance and twiss functions at 100 MeV in order to match the beam optics to beam transport line and go through it to the storage ring. To get the transverse emittance measurement, well-known technique, quadrupole scan, is used at the pre-injector. The emittance were 0.591 mm-mrad in horizontal and 0.774 mm-mrad in vertical direction.

Email:bjlee707@postech.ac.kr

Fax: +82-54-279-1401

# I. INTRODUCTION

Pohang light source-II (PLS-II) storage ring provides photon beam to beam line with high quality beam. From 2009 PLS-II Linac upgrade has been started increasing its beam energy from 2.5 GeV to 3 GeV [1]. After commissioning, 3 GeV beam was delivered to the user in March 2012. Since end of 2014, we injected the beam up to 400 mA in the top-up operation [2, 3]. Top-up operation requires high beam quality. One of the most important parameters to identify the beam quality is a transverse emittance of electron bunches. The beam emittance and Twiss functions at 100 MeV pre-injector PLS-II Linac are measured in order to match the beam optics to beam transport line and eventually inject it to the storage ring. To perform the transverse emittance measurement, well-known technique, quadrupole scan, is used at the pre-injector shown in Fig. 1. Since flexibility of focusing optics in the Linac based on energies of the order of 100 MeV, the focusing optics like triple quadrupole is used to converge beam into small angle through the transfer line. Therefore to obtain the emittance for PLS-II Linac, the quadrupole triplet and diagnostic with screen of OTR located after the quadrupole are used. A focusing (QTL) and two defocusing (QS1 and QS2) quadrupole scan are performed in the horizontal and vertical planes, respectively.

# II. EXPERIMENTAL SETUP AND QUADRUPOLE SCAN METHOD

By using the quadrupole scan technique [5, 6] for the beam emittance measurement, the triplet quadrupole consisting of single focusing (QTL) and two defocusing (QS1 and QS2) elements (see Fig. 3) is used to converge and diverge the beam at the OTR screen. The QTL, QS1 and QS2 are located at 0.830 m, 1.103 m and 0.657 m from the screen, respectively. The effective length of QTL is 0.2 m and QS1 and QS2 are 0.1 m. In the horizontal axis, only the strength of the QTL is scanned with recording the beam size in pixel unit on the OTR screen. Similarly, for the vertical axis, the QS2 is used alone. From the recorded image, we need a data processing to obtain reliable $\sigma$ of the image to calculate the beam size as function of $1/f$.

The measurement method for the beam emittance is based on the quadrupole scan technique using linear transformation matrix [4] in phase space. In general, the beam is described in two dimensional phase space leading to obtain the ellipse formalism and defined in the beam matrix with well-known beam parameters as

$$\sigma = \begin{pmatrix} \sigma_{11} & \sigma_{12} \\ \sigma_{21} & \sigma_{22} \end{pmatrix} = \varepsilon \begin{pmatrix} \beta & -\alpha \\ -\alpha & \gamma \end{pmatrix}$$

where $\sigma$ is the beam matrix and $\varepsilon$ is the beam emittance with Twiss functions ($\beta, \alpha$ and $\gamma$). The beam matrix in phase space propagated from the focusing optics to the OTR screen target is

$$\sigma_s = M \sigma_q M^T$$

with beam transformation matrix $M$ from the center of quadrupole to the OTR screen. In this case, the transformation matrix of the quadrupole is based on the thin lens approximation because the quadrupole effective length is small compared to its focal length (or the distance from the quadrupole to the screen). The beam matrix $\sigma_q$ and $\sigma_s$ are at the quadrupole and screen, respectively. To determine the beam matrix $\sigma_q$ at the quadrupole, beam size $\sigma_{s,11}$ at location of the monitoring screen has been measured and given by

$$\sigma_{s,11} = M_{11}^2 \sigma_{q,11} + 2 M_{11} M_{12} \sigma_{q,12} + M_{12}^2 \sigma_{q,22}$$

with the transformation formalism composed of thin quadrupole and drift space. The transformation matrix becomes

$$M = \begin{pmatrix} 1 & D \\ 0 & 1 \end{pmatrix} \begin{pmatrix} 1 & 0 \\ -1/f & 1 \end{pmatrix} = \begin{pmatrix} 1 - D/f & D \\ -1/f & 1 \end{pmatrix} = \begin{pmatrix} M_{11} & M_{12} \\ M_{21} & M_{22} \end{pmatrix}$$

where D is the drift from the middle of quadrupole to the screen. As this result, the beam size at the OTR screen can be written in terms of quadrupole focal length or strength as

$$\sigma_{s,11}(kl) = D^2 \sigma_{q,11}(kl)^2 + (-2D\sigma_{q,11} - 2D^2\sigma_{q,12})(kl) + (\sigma_{q,11} + 2D\sigma_{q,12} + D^2\sigma_{q,22}) \quad (1)$$

As above, the beam size is a function of quadratic of *kl*. If the beam size is fitted with a parabolic equation $(ax^2 + bx + c)$, the beam matrix at the quadrupole $\sigma_{q,011}, \sigma_{q,012}$ and $\sigma_{q,022}$ will be determined by

$$\sigma_{q,11} = \frac{a}{D^2} \quad (2)$$

$$\sigma_{q,12} = \frac{-2D\sigma_{q,11} - b}{2D^2} \quad (3)$$

$$\sigma_{q,22} = \frac{c - \sigma_{q,11} - 2D\sigma_{q,12}}{D^2} \quad (4)$$

where a, b and c are the coefficients of quadratic, linear and constant terms of the parabolic curve respectively. Subsequently, all beam parameters $(\beta, \alpha, \gamma \text{ and } \varepsilon)$ at the initial location can be determined and calculated using definition of the beam matrix at location *i* so we have

$$\varepsilon = \sqrt{\sigma_{i,11}\sigma_{i,22} - \sigma_{i,12}^2}$$

$$\beta_i = \sigma_{i,11}/\varepsilon$$

$$\alpha_i = -\sigma_{i,12}/\varepsilon \quad (5)$$

$$\gamma_i = \sigma_{i,22}/\varepsilon$$

Because the focusing properties of the quadrupole converge beam in one plane but diverge beam in the other plane, therefore to measure the beam emittance, the focusing quadrupole QTL is used in horizontal plane, while the defocusing quadrupole QS1 or QS2 is used in vertical plane. As we know, the beam is focused by the quadrupole to a focal point and it diverges again beyond the focal point as shown in Fig. 3. To measure the horizontal beam emittance the quadrupole strengths are varied while observing the beam size at an OTR screen. The image from the OTR is recorded by a CCD camera and the signal transferred to the control room.

## III. DATA ANALYSIS

The image in Fig. 4 has been project in vertical and horizontal plane in order to fit them with a Gaussian function by following

$$f(x) = A + B \exp\left(-\frac{(x-x_o)^2}{2\sigma^2}\right)$$

where A, B, $x_0$ and $\sigma$ are un-known variables. These parameters can be determined using developed MATLAB program. First the image data will be projected onto the axis which the emittance is being measured to evaluate the median over at least three points in order to remove spike. Second the ansatz for the background A can be determined by applying moving average to the data and to define the minimum. With the peak $x_0$ as the maximum of the average moving, the ansatz of the peak can be found. Fitting will be started at $B = x_0 - A$ then find the ansatz for $\sigma$ as $\left(V/\sqrt{2\pi}\right)^{1/3}$, where V is the momentum of the data, for instance data $y_i$, defined as $V = \sum_i (x-x_0)^2 . y_i$. Finally the distance between the ansatz and data has been minimized to obtain optimum set of the unknown variable A, B, $x_0$ and $\sigma$. Fig. 4 shows the data processing to fit the beam size with the Gaussian function. As this result, we found that it is fit well with the experimental data. With calibration factor of the OTR screen of 50 mm/570 pixel, we can calculate the $\sigma$ beam size in mm unit.

## IV. RESULTS AND DISCUSSIONS

For horizontal emittance measurement, the energy of the electron bunch is 98.16 MeV and the pulse length from the cathode is 2 ns. The bunch charge can be calculated using measured bunch current or voltage at BCM2. For this measurement the voltage at BCM2 is about 0.562 V. The beam emittance is measured with parameter setting in the Linac as used for good injection efficiency to the storage ring. A scan of the focusing triplet quadrupole is performed while two defocusing quadrupole are in-activated, acquiring three images for each quadrupole strength. Using the data processing, the beam size $\sigma^2 \equiv \sigma_{11}$ in eq. (1) as function of the quadrupole focal length or $kl$. With the quadratic fitting, all the coefficient a, b and c can be determined as discussed above. From eq. (1) to (4) the beam matrix at the entrance of the quadrupole are obtained. Finally, beam parameters of emittance and Twiss functions are determined. Fig. 5 shows quadratic curve of the beam size as a function of the focusing $kl$ with coefficient a = 3.421 x $10^{-6}$, b = -7.53 x $10^{-6}$ and c = 4.22 x $10^{-6}$. The resulting horizontal beam parameters at the entrance of the quadrupole can be computed and shown in Table 1.

For the vertical emittance measurement, similarly the defocusing quadrupole QS1 or QS2 is used to measure the RMS normalized emittance in the vertical axis. We scanned both quadrupole one by one without focusing quadrupole. Both QS1 and QS2 are not effective because they are located close together and the beam already small within the triplet. However the QS1 is used to scan to perform the beam width as function of $kl$ shown in Fig. 6. With the parabolic fitting, the coefficients a, b and c are 2.981 x $10^{-6}$, -4.888 x $10^{-6}$ and 2.33 x $10^{-6}$, respectively and correspond to the beam optics displayed in Table 2. From the beam size measurement as function quadrupole strength on both planes we can see clearly that the remnant fields of quadrupole triplet are more than 1.9931 1/m$^2$ strength corresponding to current of 3 A. To make the triplet quadrupole work effectively, thus, the current driven through it should be over this current.

The emittance measurement provides the Twiss function at the middle of the quadrupole. However, these parameters in horizontal and vertical axis are not at the same position thus to obtain them at the same place, we should calculate the Twiss function backward at the reference location where is beyond the injector. In this case the reference will be the position of LIFCT1 in Fig.1. Twiss functions at the LIFCT1 are calculated from the initial optics at the entrance of the quadrupole and listed in Table 3. Consequently, these optics functions here have been used to be initial values to compute these along the pre-injector and end up at the OTR screen. As result of the Twiss function at the OTR monitoring, beam size in both directions can be determined and compared to the measured beam size. Fig. 7 shows the beam size in horizontal and vertical axis obtained from the measurement compared to the analysis at the screen.

The second defocusing quadrupole (QS1) is located too close to QS2 to allow the beam to be enlarged at QS2 for better beam emittance measurement. As a result, the defocusing quadrupole (QS2) cannot effectively create a beam waist vertical axis ahead of the screen. Therefore the experimental setup in this case does not allow the scanning of a parabolic beam height at the screen like in the horizontal case. Therefore the Twiss functions computed from this measurement are not correct and give wrong results in the test as shown in Fig.7(b). Further measurement for the vertical beam emittance with the good injection condition should be performed at the end of Linac or somewhere consisting of focusing optics which is at optimum location.

## V. CONCLUSION

Transverse beam emittance measurement has been performed at the end of pre-injector using quadrupole scan. With the setup as previous discussion, the horizontal beam emittance and Twiss parameters at the end of injector are in good agreement with simulation which is obtained from the measured beam size at the screen and satisfy the beam transport line requirement while in the vertical axis, these optics parameters are not satisfied due to convergence beam profile in the triplet. With the Twiss functions at the end of injector, we can calculate them through the beam transport line to match these parameters to the storage ring.

Table 1. Beam optics at entrance of focusing quadrupole QTL at pre-injector

| Beam parameters | value |
|---|---|
| $\varepsilon_x$ [mm-mrad] | 0.5911 |
| $\beta_x$ [m] | 6.691 |
| $\alpha_x$ | -0.1692 |
| $\gamma_x$ | 0.1537 |

Table 2. Beam optics at entrance of defocusing quadrupole QS1 at pre-injector

| Beam parameters | value |
|---|---|
| $\varepsilon_y$ [mm-mrad] | 0.774 |
| $\beta_y$ [m] | 2.897 |
| $\alpha_y$ | 0.1375 |
| $\gamma_y$ | 0.3517 |

Table 3. Twiss functions at LIFCT1 (reference location)

| Twiss functions | Horizontal | Vertical |
|---|---|---|
| $\beta$ [m] | 6.566 | 2.966 |
| $\alpha$ | 0.096 | 0.208 |
| $\gamma$ | 0.1537 | 0.1375 |

Fig. 1. PLS-II Linac layout with position of beam diagnostic station and of triple quadrupole used in transverse emittance measurement.

Fig. 2. Ray tracing through a focusing quadrupole.

Fig. 3. Schematic layout of quadrupole scanning setup for beam emittance measurement.

Fig. 4. Projection of the observed vertical beam size (squares) and Gaussian fit (solid line).

Fig. 5. Focusing quadrupole scan for RMS normalized emittance in horizontal plane.

Fig.6. Defocusing quadrupole scan for RMS normalized emittance in vertical plane.

Fig.7: Comparison of beam size between measurement and analysis obtained from measured Twiss functions in horizontal (a) and vertical (b) directions.

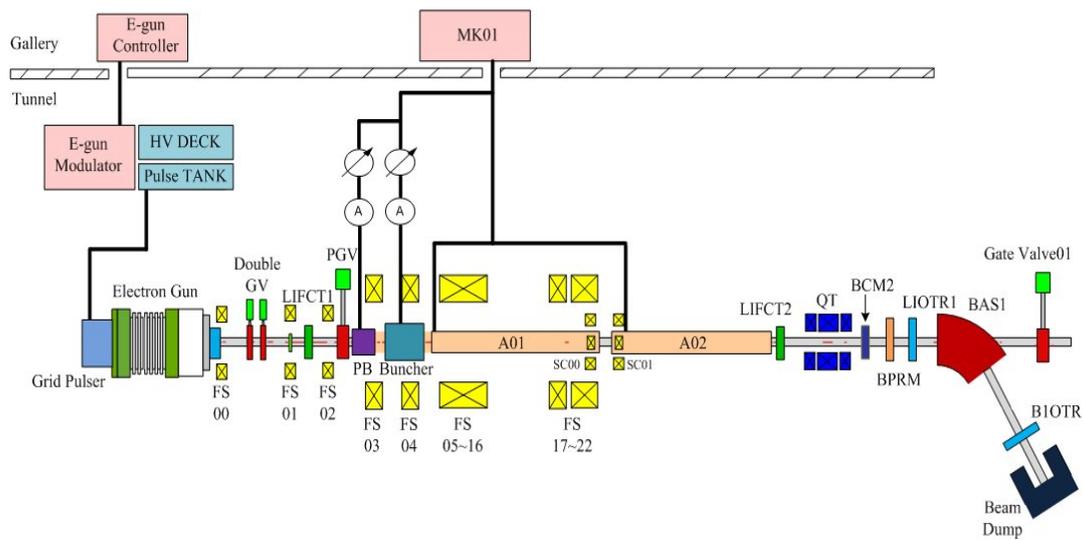

Fig. 1.

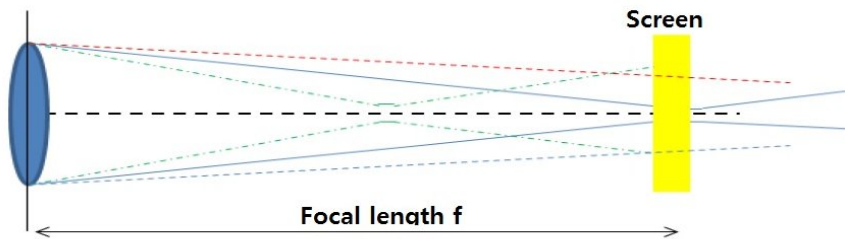

Fig. 2

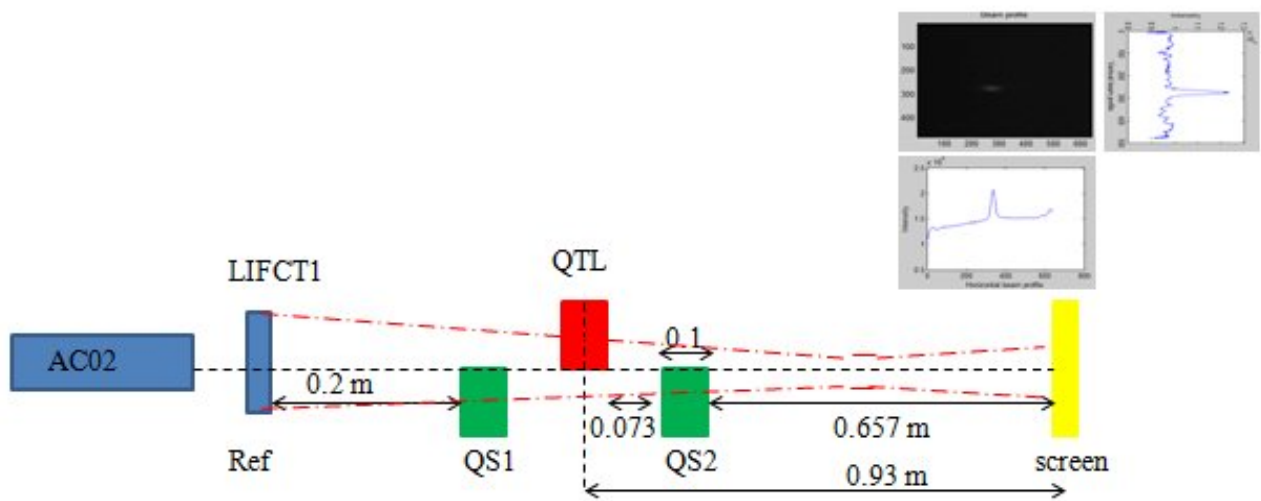

Fig. 3

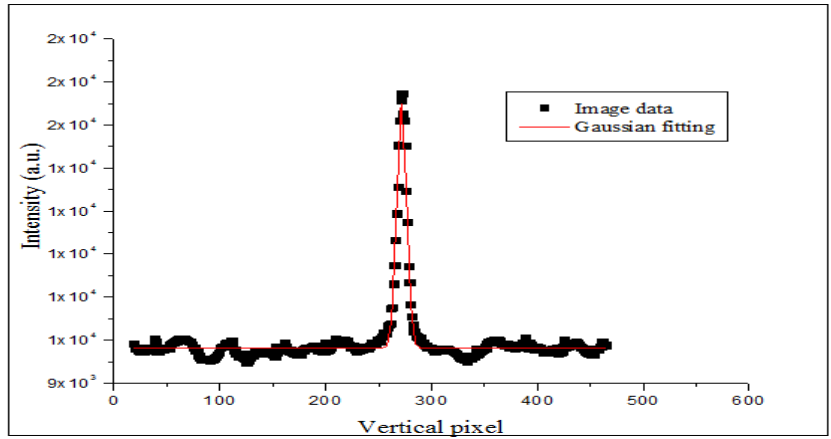

Fig. 4

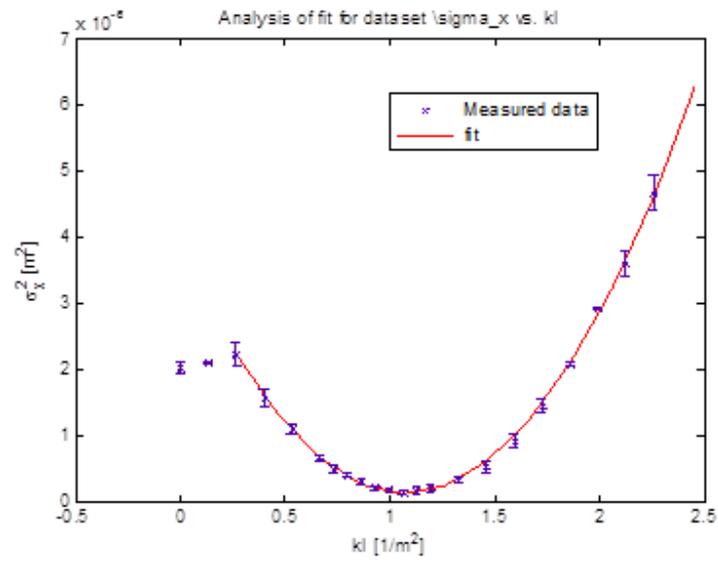

Fig. 5

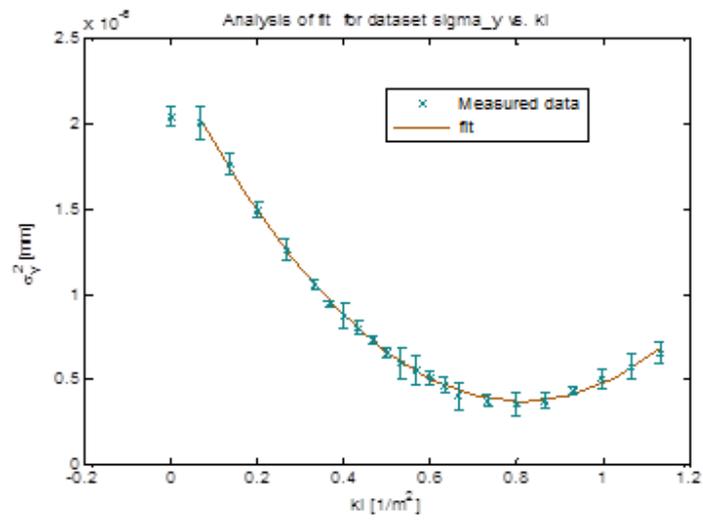

Fig. 6

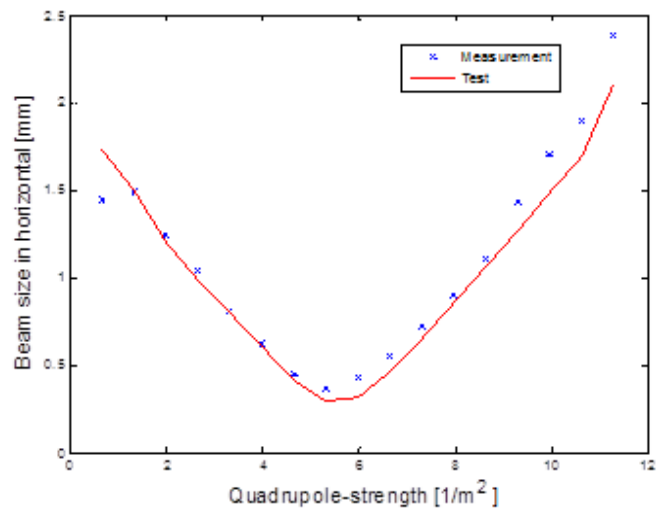

Fig. 7 (a)

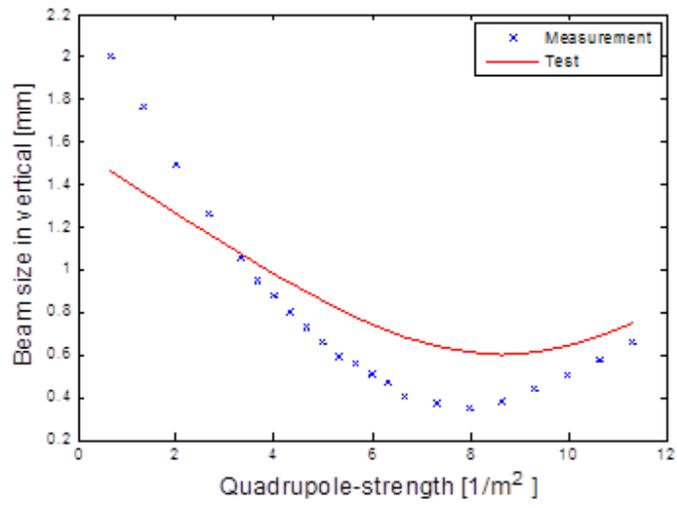

Fig. 7 (b)